\documentstyle[epsf]{article}
\newcommand{\ee}{\end{equation}}
\newcommand{\be}{\begin{equation}}
\newcommand{\eea}{\end{eqnarray}}
\newcommand{\bea}{\begin{eqnarray}}
\newcommand{\ea}{\end{array}}
\newcommand{\ba}{\begin{array}}
\begin{document}

\begin{center}
{\bf
On the Second  Law of thermodynamics and the piston problem}
\vskip 3mm
{\rm Christian Gruber and S\'everine Pache}
\vskip 1mm
{\it Institut de Th\'{e}orie
des Ph\'enom\`enes Physiques, Ecole Polytechnique
F\'ed\'erale de Lausanne, CH-1015 Lausanne, Switzerland}
\vskip 3mm
{\rm Annick Lesne} 
\vskip 1mm
{\it Laboratoire de Physique Th\'eorique 
des Liquides, 
Universit\'e Pierre et Marie Curie,  
Case  121,\\ 4 Place Jussieu, 75252
 Paris Cedex 05,
France\vskip 3mm}
\date{\today}
\end{center}

\vskip 10mm
\begin{abstract}
{\small The piston problem is investigated in the case 
where the length of the cylinder is infinite 
(on both sides)
and the ratio $m/M$ is a very small parameter, where $m$ is the mass of one particle of the gaz
and $M$ is the mass of the piston. Introducing initial conditions such that the stochastic motion
of the piston remains in the average at the origin (no drift), it is shown
that the time evolution of the  fluids,
analytically derived from Liouville equation, agrees with the Second Law
of thermodynamics.
 We thus have a non
equilibrium microscopical model whose evolution 
can be explicitly shown to
obey the two laws of thermodynamics.
}
\end{abstract}

\section{Introduction}

\indent\indent Recently the well-known ``adiabatic piston problem" has
 attracted a lot of attention  \cite{gruber}-\cite{callen}.
 Most
investigations have however concentrated on the motion of the piston and very few studies have
been made concerning the time evolution of the fluids on the two sides
 of the piston. In our
previous works \cite{GPL, gp} it was shown that the piston 
in a cylinder of finite length evolves with two different time scales. In a first time scale the
motion of the piston can be considered as ``deterministic" and ``adiabatic", and the system evolves
to a state of mechanical equilibrium where the pressures on both sides are approximately equal,
but the temperatures different. Then on a much larger
 time scale  (if $m\ll M$), the piston evolves ``stochastically"
and with ``heat transfer" to a state of thermal equilibrium, where the temperatures (and the
pressures) on both sides of the piston are equal. It was shown in [5 a]
 that in the first time
scale the  relaxational motion of the piston can be either weakly damped or strongly 
damped depending on whether
the parameter $R = M_{gas}/M_{piston}$ is small ($R<1$) or large ($R>1$) 
where $M_{gas}$ is the total mass of the gas.

\indent A microscopical analysis of this model was given in [5 b].
 We were able to derive equations describing
the time evolution based on three ``reasonable" assumptions. It was also shown that the results
obtained from these equations were qualitatively in good agreement with those observed on a very
large number of  numerical simulations, except for the observed damping coefficient which appears
a lot smaller than the predicted value. It was realized that the underlying damping mechanism
must be related to the propagation of sound waves bouncing back and forth
 between the boundaries  of
the cylinder and the piston. Unfortunaly the effect of such a wave propagation could not be taken
into account in our previous analysis, because of the ``average assumption"  we had introduced
(i.e. the values of density, pressure, temperature on the right/left surface of the piston can be
replaced by the average of these quantities in the right/left compartment). Therefore to
understand how the damping effect can be generated by purely elastic collisions
 on the piston surface and cylinder walls, we are forced to
investigate the propagation of wave through   the ideal
 and out-of-equilibrium fluids.

\indent To simplify the following investigation on the wave propagation in the fluids, we shall consider
the case where the length of the cylinder is infinite 
 on both sides and all velocities (of the particles and the
piston) are parallel to the axis of the cylinder. Initially the fluids
 on the left
and on the right of the piston are each 
in thermal equilibrium with densities $n_0^{\pm}$, temperatures
$T_0^{\pm} \,\,(T_0^+ > T_0^-)$	and pressures $p_0^{\pm}$, where
 the index $+/-$ refer to the right/left
of the piston. To  simplify
 the problem further we have choosen the pressures $p_0^{\pm}$
 (given $T_0^{\pm}$) in such a manner that
the piston, which evolves stochastically under the 
elastic collisions with the particles, remains on the
average at the origin (for a time sufficiently short 
to neglect the Brownian deviation -- a second-order
 contribution to the piston motion).
 With these initial conditions no work will be done by the right side on the left side and
there will be only heat transfer (from the right to
 the left since $T_0^+ > T_0^-)$. Following the
analysis  in [5 b], which is a  singular
perturbative approach in   powers of the small parameter $m/M$, we shall
derive, for these initial conditions, the distribution function for the velocity of the piston, in
the stationary state (of the piston), to first order in $\sqrt{m/M}$. Using the fact that we have  only
elastic collisions in our model, we shall then derive the distribution function $\rho(x,v;t)$ for the
fluids. Finally we shall obtain from the well-known relations of kinetic theory the density fields
$n(x,t)$, the velocity fields $w(x,t)$, the stress (tensor) $\tau(x,t)$ and the ``heat" flux
$j_Q(x,t)$. In other words we will obtain the time evolution of our microscopical model,
considered as a fluid out of equilibrium.

\indent The next problem we shall discuss
 in the following is whether the evolution thus obtained obeys
the laws of non-equilibrium thermodynamics. In particular the question will be whether it is
possible to find  a microscopically rooted
entropy function $s$ from which, following the laws of thermodynamics, we
recover the evolution of the fluids.

It will be shown that is possible to consider our fluid as a two-component simple fluid described
by a fondamental equation $s=s(u,n_1,n_2)$ from which we can recover our fluid equations, together
with the well-known phenomenological equations of thermodynamics, and an entropy production which is
strictly positive. Moreover this entropy function will coincide 
 with the Boltzmann  entropy at
those points where the fluid is in a stationary state (i.e. at $\pm \infty$ and on the right/left
surface of the piston).

In Section~2 we shall recall the thermodynamic equations of the 1-component and the 2-component
simple fluids. We then recall in Section~3 the microscopical definitions introduced in kinetic
theory for the thermodynamical variables. The piston problem  and the distribution function for
the velocity of the piston in the stationary state are presented in Section~4. We then discuss the
properties of the fluids at the surface of the piston in Section~5, and the general evolution of
the fields in Section~6. The problem of entropy and entropy production is discussed in
Section~7-8 and  numerical
graphs of the different fields are  presented for a specified set of initial conditions.  Finally, general conclusions are presented in the last section.


\section{Thermodynamics of 1-dimensional simple fluids}

It is well known that the conservation of the number of
particles, the linear momentum, and the energy, i.e. the First
Law of thermodynamics, give for a single component fluid in 1  dimension
the following equations for the density of particles
$n(x,t)$,  the fluid velocity  $w(x,t)$
and the energy density $e(x,t)$:
\bea
&& \partial_tn+\partial_x(nw)=0\label{eq1}\\
&&mn[\partial_tw+w\partial_xw]-\partial_x\tau=0\label{eq3}\\
&&\partial_te+\partial_x(ew+j_e)=0\label{eq4}
\eea
where
$m$ is the mass of one particle, $\tau$ is the stress
(tensor) and $j_e$ is the energy current,
 adding to the ``convective" contribution $ew$. 
We have also assumed that the fluid is not submitted to
external forces. 
Furthermore, the Second Law of thermodynamics yields for the
entropy density $s(x,t)$:
\be\label{eq4bis}
\left\{\ba{l}
\partial_ts+\partial_x(sw+j_S)=i\\
\\
i=i(x,t)\geq 0  \hskip 5mm \mbox{\rm for all}\;\;(x,t)
\ea\right.\ee
where $j_S(x,t)$ is the entropy current and $i(x,t)$ the
``internal
entropy production" or ``irreversibility". 

A simple fluid is defined by the condition that there exists a
function $u=u(s,n)$,  independent of $x$ and $t$, 
 such that:
\be
e={1\over 2}\;mnw^2+u(s,n)\ee
and in this case, one introduces the  temperature
$T(x,t)$ and the pressure $p(x,t)$ by the thermodynamical
definitions:
\be\label{eq6}
T={\partial u\over \partial s}, \hskip 10mm
\mu={\partial u\over \partial n},\hskip 10mm
p=Ts+\mu n-u\ee
It follows from Eqs.~(\ref{eq1}-\ref{eq6}) that:
\bea
&&j_e=-\tau w +j_Q\\
&&j_Q=Tj_S\label{eq8}\\
&&i=j_Q\partial_x\left({1\over T}\right)
+{1\over T}(\tau+p)\partial_xw \label{eq9}\eea

The description  of a two-component fluid involves 
an additional equation, for the density 
$n_1(x,t)$ of the first component
 and the associated current $j_1(x,t)$:
\be\label{eq10bis}
 \partial_tn_1+\partial_x(n_1w+j_1)=0\label{eq1prime}\ee
Assuming that the particles of the two components
have all the same mass, the simple fluid is now defined by the
condition that
 there exists a
function $u=u(s,n_1,n_2)$ such that:
\be
e={1\over 2}\;mnw^2+u(s,n_1,n_2),\qquad
{\rm with}\;\;\; n = n_1 + n_2\ee
Again one introduces the  temperature
$T(x,t)$ and the pressure $p(x,t)$ by the thermodynamical
definitions:
\be\label{eq6prime}
T={\partial u\over \partial s}, \hskip 10mm
\mu_i={\partial u\over \partial n_i},\hskip 10mm
p=Ts+\mu_1 n_1+\mu_2 n_2-u\ee
In that case we obtain
\bea
&&j_e=-\tau w +j_Q\label{jq}\\\
&&j_Q=Tj_S+j_1(\mu_1-\mu_2)\label{eq8prime}\label{jq2}\\
&&i=j_Q\partial_x\left({1\over T}\right)
+{1\over T}(\tau+p)\partial_xw +j_1\;
\partial_x\left({\mu_2-\mu_1\over T}\right)\label{eq15bis}\eea
Let us stress that the ``heat current" is not a uniquely defined concept in the case of a
two-component fluid. On can either introduce the heat current by the condition that the sum
of the ``heat" current
plus the ``work" current $-\tau w$
 yields the ``energy" current $j_e$, 
 so that  $j_Q$ defined above by Eq.~(\ref{jq})
 is interpreted as the heat current. Alternatively 
one can define a heat current $j_H$
by the condition that Eq.~(\ref{eq8}), 
 written above for a one-component fluid, 
remains the phenomenological relation between the ``heat"
and the ``entropy" currents, i.e.
\bea
&&j_H=Tj_S=j_Q  - j_1(\mu_1-\mu_2)  \label{jh}\\\
\eea With this definition the entropy production is expressed by 
\bea
&&i=j_H\partial_x\left({1\over T}\right)
+{1\over T}(\tau+p)\partial_xw +\frac{1}{T} \;j_1\;
\partial_x\left({\mu_2-\mu_1}\right)\label{eq15-2}\eea which 
is the usual form in thermodynamics.

The irreversibility  $i(x,t)$  is thus  a sum of products [``current"
times ``force"] and usually one introduces at this point
phenomenological relations between forces and currents to 
ensure that the irreversibility
is non-negative for all $(x,t)$.
Indeed, whereas the First Law Eqs.~(\ref{eq1}-\ref{eq4})
can be understood from any microscopic viewpoint
and has the status of an exact and inviolable necessity, the
Second Law Eq.~(\ref{eq4bis}), which is of fundamental importance for non-equilibrium 
macroscopical physics, is a statement concerning concepts, "entropy" and 
"entropy production", for which there seems to be no agreement on a microscopical 
definition, except for equilibrium states, and  whose necessity for non-equilibrium 
situations is even subject to controversy \cite{Gal, lieb1}.

In the following we shall consider this issue for the piston problem described in the introduction. For this problem we are led to consider
 the { special case}  where 
\be\label{eq10}
u={1\over 2}\,nk_BT=-\,{\tau\over 2}\ee
It then follows that:
\be\label{2c}
s(u,n_1,n_2)=
nk_B\left[{1\over 2}\left(
1+\ln{4\pi u\over mn^3}\right)+g(n_1,n_2)\right]\ee
where $g=g(n_1,n_2)$ is so far undetermined and:
\be
p=nk_BT\left(1-n_1{\partial g\over \partial n_1}
-n_2{\partial g\over \partial n_2}\right)
\ee
In that case the viscous-stress (tensor) 
$\tau_{fr}$ is given by:
\be
\tau_{fr}=\tau+p=-\, nk_BT\left(n_1{\partial g\over \partial n_1}
+n_2{\partial g\over \partial n_2}\right)
\ee
Note that if the two fluids are identical and in equilibrium, then using
the definition of  Boltzmann entropy (Section~3)
 for $s(x,t)$, we have
$g(n_1,n_2)\equiv 0$.

\section{Kinetic theory}

 To obtain a microscopical definition of the entropy density 
$s(x,t)$, we consider the kinetic framework underlying the above
 fluid description.
We consider a system of point particles, with mass $m$, in a
semi-infinite cylinder $x\leq 0$.
We assume that all velocities are parallel to the axis 
of the cylinder (i.e. the $x$-axis) and that the particles
interact through pure elastic collisions only.
 In kinetic theory,
this ``one-dimensional" system is described by a distribution
function $\rho(x,v,t)$ solution of 
the  continuity equation:
\be
(\partial_t+v\partial_x) \rho(x,v,t)=0
\hskip 10mm{\rm for}\;\;x<0\ee
together with the boundary condition at $x=0$:
\be\int_{-\infty}^{+\infty}
dv\;\rho(x,v,t)v=0\ee
It is well known that one recovers the conservation equations
for the one-component fluid Eqs.~(\ref{eq1}-\ref{eq4})
by means of the following definitions:
\bea
&&n(x,t)=\int_{-\infty}^{+\infty}
dv\;\rho(x,v,t)\\
&&w(x,t)={1\over n(x,t)}\;\int_{-\infty}^{+\infty}
dv\;\rho(x,v,t)v\\
&&\tau(x,t)=-\,m\;\int_{-\infty}^{+\infty}
dv\;\rho(x,v,t)(v-w(x,t))^2\\
&&j_Q(x,t)={m\over 2}\;\int_{-\infty}^{+\infty}
dv\;\rho(x,v,t)(v-w(x,t))^3\label{eq21}\\
&&e={1\over 2}\,nmw^2+u\\
&&u=-\,{1\over 2}\,\tau\\
&&j_e=-\tau w+j_Q
\eea
We would like to complete these equations with a continuity
equation for the entropy.
In other words we want to define the fields 
entropy density, current of entropy, production of entropy, as well as the temperature and pressure fields.

 One can introduce for example
the {\it  Boltzmann entropy}:
\be\label{29}
s_B(x,t)=-k_B\; \int_{-\infty}^{+\infty}
dv\;\rho(x,v,t)\ln \rho(x,v,t)\label{gib}\ee
which obeys the equation:
\be\label{eq25}
\partial_ts_B+\partial_x(s_Bw+j_B)=0\ee
with: 
\be\label{eq26}
j_B(x,t)=-k_B\; \int_{-\infty}^{+\infty}
dv\;\rho(x,v,t)[v-w(x,t)]\ln \rho(x,v,t)\ee
However we do not have a standard definition
of the ``temperature". If we identify $j_B$
with the current of $s_B$, then to obtain the 
fluid equations (\ref{eq4bis})  and  (\ref{eq8}), we are led to
define:
\be\label{eqTB}
T_B(x,t)={j_Q(x,t)\over j_B(x,t)}\hskip 10mm{\rm and}
\hskip 10mm i_B(x,t)=0\ee
The condition $i_B=0$ is understood from kinetic theory where the
internal entropy production is  induced by the collision term
which is zero in our case. The problem with the above definition of
temperature is that we do not know whether  thermodynamic
relations between $n$, $u$, $s_B$, $T_B$ such as  Eq.~(\ref{eq6})  will be
satisfied.

\vskip 3mm
If on the other hand we adopt the usual definition of
temperature as related to the thermal energy, i.e. for non interacting
particles in 1 dimension:
\be\label{eq30}
u={1\over  2}\;nk_BT\ee
and the relation:
\be
j_Q=Tj_S\ee
then we have:
\be
\partial_ts_B+\partial_x\left(s_Bw+{j_Q\over T}\right)=i_B\ee
with $i_B(x,t)$ given by:
\be i_B=\partial_x\left({j_Q\over T}-j_B\right)\ee
but  we cannot conclude at this point that $i_B(x,t)$
will be non-negative.
Furthermore we do not have in general the
thermodynamic relation $1/T = \partial s_B/\partial u$.

\vskip 3mm


 We could also consider that the fluid defined by $\rho(x,v,t)$ is a
simple 1-component fluid with temperature given by the thermal temperature  Eq.~(\ref{eq30}). In that
case we have for the thermodynamic entropy, pressure, and viscous
stress:
\be\label{eq31}
s(x,t)=
nk_B\left[{1\over 2}\left(
1+\ln{4\pi u\over mn^3}\right)+g(n)\right]\ee
\be\label{eq32}
p=nk_BT-n^2k_BTg^{\prime}(n), \hskip 10mm {\rm hence}\;\;\; \tau_{fr}=\tau+p=
-n^2k_BTg^{\prime}(n)\ee
We would have to find at this point  an appropriate function $g(n)$ such that
$ t\,i(x,t)$, Eq.~(\ref{eq9}), is non negative. The simplest choice would be  $g= 0$ in order
that $s$ coincide with $s_B$ at equilibrium.

\vskip 3mm

 In the example  presented in the introduction
(piston problem, see also Section~4) the entropy at
 point $(x,t)$ must depend on 3 parameters $(n_0,T_0,T_P)$
 where $n_0$, $T_0$
are the density and temperature at $x=-\infty$
 and $T_P$ the piston
temperature. It is thus not possible to consider the fluid as a simple
one-component fluid described by 2 state variables, i.e. $s =s(u,n)$.
However,  we shall see that it is possible to consider the
system as a 2-component fluid with an entropy function:
\be
s(x,t)=s[u(x,t), n_1(x,t),n_2(x,t)]
\ee
which coincide with the Boltzmann  entropy at $x=-\infty$  
and $x=- 0$, where the fluid is in a stationary state.
 In that
case we are led to introduce a temperature $T_s$  by the thermodynamic
definition:
\be
{1\over T_s}={\partial \over \partial u}\;s(u,n_1,n_2)\ee
a current $j_1$
 satisfying the continuity equation Eq.~(\ref{eq10bis})  and an
entropy flux now given by Eq.~(\ref{eq8prime}). Again with this choice we have to
verify whether $i(x,t)$ given by Eq.~(\ref{eq15bis})  is 
 actually non negative.

\vskip 3mm


The purpose of the following sections is to explore what these
different definitions imply for the piston problem, and whether
we can arrive for this model at a description consistent with the 
thermodynamics  of fluids, i.e. the two laws of thermodynamics.

\section{Piston problem}

We consider an infinite cylinder of area $A$ containing two
gases separated by a movable piston. 
The gases are made of point
particles with mass $m$, while the piston is a rigid solid 
with mass  $M\gg m$ and no internal degrees of freedom.
The particles and the piston interact through purely elastic
collisions only.  We assume that all velocities are parallel to
the $x$-axis (the axis of the cylinder) which reduces
the system to one dimension. Initially the piston is at rest and
the two gases are homogeneous and in thermal equilibrium
described by Maxwellian distributions of velocities with
temperatures $T_0^{\pm}$,
hereafter denoted $\varphi_{T_0^{\pm}}$,  and densities $n_0^{\pm}$
respectively on the left $(-)$ and on the right $(+)$
of the piston. 
We shall  take initial conditions $n_0^{\pm}$, $T_0^{\pm}$
such that the piston evolves toward an equilibrium
state
with $\langle V\rangle =0$,  and  for clarity we choose
  $T_0^+>T_0^-$.
Using the results of [5 b],
this requires that the initial pressures
$p_0^{\pm}=n_0^{\pm}k_BT_0^{\pm}$ satisfy the condition:
\be\label{eq35} p_0^+-p_0^-={m\over
2M}(p_0^++p_0^-)\left(\sqrt{T_0^+\over T_0^-}-
\sqrt{T_0^-\over T_0^+}\right)+{\cal O}\left[
\left({m\over M}\right)^2\right]\;>0\ee
Using a  singular perturbative approach to first order in $m/M$, it was
shown that the piston will reach an equilibrium state
$\Phi(V)$, with $\langle
V\rangle=0$ and  temperature $T_P$
where
\be\label{eq36}
k_BT_P \equiv M \langle
V^2\rangle =k_B\sqrt{T_0^+T_0^-}\ee
in a time $t_1$ which is proportional to $M/A$
and can be made arbitrary small.
In this equilibrium state of the piston, the heat passing through
the piston per unit time is given by [5 b]:
\be\label{eq37}
P_Q^{(+)\to(-)}={Am\over 2M}\;
\sqrt{8k_B\over m\pi}\;(p_0^++p_0^-)
\left(\sqrt{T_0^+}-\sqrt{T_0^-}\right)+
{\cal O}\left[
\left({m\over M}\right)^2\right]\;>0\ee
In the present paper,
we take this equilibrium state of the piston and
$X = 0$ as the initial conditions for the piston and  Maxwellian
distribution with $T^{\pm}_0$
for the gases.
The piston will henceforth diffuse around $X=0$, but this
diffusive motion is very slow
compared to the relaxation phenomena inside the gases (damping
oscillations) 
 and it can be neglected on the time scale 
here considered.

\vskip 1mm
In  \cite{GP}, the velocity distribution function
of the piston $\Phi(V)$ in the stationary state was obtained to
first order in $\sqrt{m/M}$ for the case $p_0^+=p_0^-$. The same 
analysis was conducted for the case  $p_0^+\neq p_0^-$
in \cite{dip}. For our considerations where $p_0^+$ and $p_0^-$
satisfy the condition Eq.~(\ref{eq35}), one obtains:
\be\label{eq39}
\Phi(V)\sim e^{-\beta_PV^2}\left\{
1+{1\over 2}\sqrt{m/M}\;a_1\sqrt{\beta_P}\;V-{1\over 3}
\sqrt{m/M}\;a_1\left(\sqrt{\beta_P}\;V\right)^3+{\cal O}
\left({m\over M}\right)
\right\}
\ee
with
\be
\beta_P={M\over 2k_B\sqrt{T_0^+T_0^-}}=
{M\over 2k_BT_P}, \hskip 10mm
a_1=\sqrt{\pi}\left(\eta-{1\over \eta}\right), \hskip 10mm
\eta=\left(T_0^+\over T_0^-\right)^{1/4}
\ee
Assuming that we can neglect the recollisions of particles on
the piston, or better, introducing a cutoff in the velocity
distribution such as was done in \cite{Cher}-\cite{Cher3},  \cite{Pi}-\cite{LPS2} to ensure that this
condition is satisfied, one can thus compute the velocity
distribution of the particles at the surface of the piston.
The
distribution of velocities of the particles before 
($P(v)$) and after ($\widetilde{P}(v)$) a
collision on the piston are given respectively by:
\be\label{eq39bis}
P^-(v)\sim\int_{-\infty}^vdV\varphi_{T_0^-}(v)\Phi(V)
\hskip 10mm{\rm and}\hskip 10mm
\widetilde{P}^-(v)\sim \int_v^{\infty}
dV\varphi_{T_0^-}(v^{\prime})\Phi(V^{\prime})
\ee
on the left side and
\be\label{eq39ter}
P^+(v)\sim \int_v^{\infty}
dV\varphi_{T_0^+}(v^{\prime})\Phi(V^{\prime})
\hskip 10mm{\rm and}\hskip 10mm
\widetilde{P}^+(v)\sim \int_{-\infty}^vdV\varphi_{T_0^+}(v)\Phi(V)\ee
on the right  side
where:
\be
\left\{
\ba{l}
v^{\prime}=-(1-\alpha)v+(2-\alpha)V\\
\\
V^{\prime}=(1-\alpha)V+\alpha v
\ea
\right.\qquad
\alpha={2m\over M+m}\ll 1\ee
Introducing the boundary condition at the surface of the piston

\be
\int_{-\infty}^{+\infty}dv \rho^{\pm}(0,v,t)v=0\ee
 and using the fact that $\Phi(V)$ is peaked
in $V=0$,
of width $(k_BT_P/M)^{1/2}$ far smaller than typical $v$,
we  obtain  the distribution function of the
particles on the left surface of the piston:
\be\label{eq43}
\rho^{-}(0,v,t)=
\theta(v)\rho^{-}(v)+\theta(-v)\widetilde{\rho}^-(v)
\ee
where
\be\label{eq56}
\left\{
\ba{l}
\rho^{-}(v)=n_0^-\;\displaystyle
\sqrt{\beta^-\over \pi}\;e^{-\beta^-v^2}\\
\\
\widetilde{\rho}^-(v)=\displaystyle{1\over 1+\delta^-}\;n_0^-
\;\sqrt{\beta^-\over \pi}\;e^{-\widetilde{\beta}^-v^2}
\ea
\right.\ee
with
\be\label{eq46}
\left\{
\ba{l}
\beta^-=\displaystyle{m\over 2 k_BT_0^-}\\
\\
\widetilde{\beta}^-={\displaystyle\frac{\beta^-}{1+\delta^{-}}}\\
\\
\displaystyle\delta^{-}=\alpha(2-\alpha)\left({T_P\over T_0^-}-1\right)=
{\cal O}
\left({m\over M}\right)>0
\ea
\right.\ee
Similarly  the distribution function on the right
surface of the piston is given by:
\be\label{eq47}
\rho^{+}(0,v,t)=
\theta(-v)\rho^{+}(v)+\theta(v)\widetilde{\rho}^+(v)
\ee
 together with Eq.~(\ref{eq56}) where $\beta^-$, 
$\widetilde{\beta}^-$ and $\delta^-$ 
are replaced
by
\be
\left\{
\ba{l}
\beta^+=\displaystyle{m\over 2 k_BT_0^+}\\
\\
\widetilde{\beta}^+={\displaystyle\frac{\beta^+}{1+\delta^{+}}}\\
\\\displaystyle
\delta^{+}=\alpha(2-\alpha)\left({T_P\over T_0^+}-1\right)=
{\cal O}
\left({m\over M}\right)<0
\ea
\right.\ee
Strictly speaking the distributions (\ref{eq43})
and  (\ref{eq47})  are valid only for $|v|\gg (k_BT_P/M)^{1/2}$,
because of the error functions which appear in
(\ref{eq39bis}) and (\ref{eq39ter}).
{
Indeed, in  (\ref{eq39bis}) and (\ref{eq39ter},
we replaced $\int_{-\infty}^v dV$ by $\theta[v]\,\,\int_{-\infty}^{+\infty}dV$
, and $\int_v^{+\infty} dV$
 by $\theta[-v]\,\,\int_{-\infty}^{+\infty}dV$ , which is
valid only if $|v|$ is far larger than the width of $\Phi(V)$,
namely $|v|\gg\sqrt{\langle V^2\rangle}= (k_BT_P/M)^{1/2}$.
}
However in the following,
we shall
be interested only in  thermodynamic quantities defined by
averages with respect to $\rho^{\pm}(0,v,t)$ and since $M\gg m=1$
(typically  $M=10^3-10^6 $ in our simulations),
the corrections due to error functions will be negligible.


\section{Thermodynamical quantities at the surface of the piston}

Given the distributions (\ref{eq43})  and (\ref{eq47}), the
thermodynamical
quantities defined in Section~3 are immediately obtained at the
surface of the piston ($x=0$). On the left side we have:
\be
n^-(0,t)={1\over 2}\;n_0^- \left(1+{1\over
\sqrt{1+\delta^-}}\right)\ee
\be\label{58}
 2u^-(0,t)=-\;\tau^-(0,t)=
{1\over 2}\;p_0^-\left(1+\sqrt{1+\delta^-}\right)=
p_0^-\left(1+{m\over M}\left({T_P\over
T_0^-}-1\right)\right)+{\cal O}\left[
\left({m\over M}\right)^2\right]\ee
\be\label{59}
T^-(0,t) = {2u^-(0,t)\over k_Bn^-(0,t)} = T^-_0\sqrt{1+ \delta^-}\ee
\be\label{eq52}
j_Q^{-}=-\;{1\over
2}\;\delta^{-}p_0^{-}\;\sqrt{2k_BT_0^{-}\over m\pi}\;<0\ee
and similarly for the right side,
 with corresponding quantities $n_0^+$, $p_0^+$,
$T_0^+$ and $\delta^+$,  except for
 $j_Q$ where an additional sign change is involved:
\be\label{eq52+}
j_Q^{+}=+\;{1\over
2}\;\delta^{+}p_0^{+}\;\sqrt{2k_BT_0^{+}\over m\pi}\,\, <\,\,0\ee
As can be seen from (\ref{58}), the 
 (mechanical) equilibrium condition of the piston, 
i.e. the equality
of the forces on both side (since $\langle V\rangle =0$):
\be
\tau^-(0,t)=\tau^+(0,t)\hskip 10mm\mbox{\rm for all}\;\;t\ee
is to first order in $m/M$ identical
with the condition Eq.~(\ref{eq35}) previously obtained at this
order  (in the framework of kinetic theory).
Moreover, from the stationarity condition expressed by
\be
j_Q^-(0,t)=j_Q^+(0,t)=j_Q(0,t)
\hskip 10mm\mbox{\rm for all}\;\;t\ee
we recover at lowest order in $m/M$ our previous results 
Eqs.~(\ref{eq36}-\ref{eq37}), i.e.
\bea
T_P&=&\sqrt{T_0^+T_0^-}+{\cal O}\left({m\over M}\right)
\hskip 10mm ({\rm hence}\;\;\; T_0^-<T_P<T_0^+)
\\
j_Q&=&-\,{1\over A}P_Q^{(+)\to(-)}+{\cal O}\left[
\left({m\over M}\right)^2\right]\;<0\label{jqq}
\eea
For the  Boltzmann  entropy Eqs.~(\ref{eq43})  and (\ref{gib}) give:
\be\label{eq53}
\left\{
\ba{l}\displaystyle
s_B(0,t)=n(0,t)k_B\left\{
{1\over 2}\left(1+\ln {4\pi\over m}\right)+{1\over 2}
\ln {u(0,t)\over n^3(0,t)}+C(\delta)
\right\}\\\displaystyle
\\\displaystyle
C(\delta)=\left(
{1\over{1+\sqrt{1+\delta}}}-{3\over 4}\right)\ln(1+\delta)
+\ln\left(1+\sqrt{1+\delta}\right)-\ln 2
=-\,{3\over 32}\delta^2+{\cal O}\left[
\left({m\over M}\right)^3\right]
\ea
\right.\ee
where the subscript $(+/-)$ for right/left have been omitted
and the currents $j_B^{\pm}$ of Boltzmann entropy (\ref{eq26})  are
\be\label{eq56-2}
j_B^{\pm}(0,t)=\,\,{\pm}\,\,\,{1\over 2}\,\ln(1+\delta^{\pm})\,\label{jjg}
n_0^{\pm}k_B\,\sqrt{2k_BT_0^{\pm}\over m\pi}\;<0\ee
Finally integrating the continuity equation (\ref{eq25}), the  Boltzmann entropy $ S^{\pm}_B$ on the right/left compartment satisfies equation :
\be\label{e68}\label{esg}
{d\over dt} S_B^{\pm}(t)=\pm\; A\;{j_Q(0,t)\over T^{\pm}(0,t)}
\;\sqrt{1+\delta}\;
{\ln(1+ \delta)\over \delta}  \qquad (\delta = \delta^{\pm})\ee
From (\ref{jqq}, \ref{jjg}), 
 the ``Boltzmann temperature" (\ref{eqTB}) is:
\be
  T_B(0,t) = T(0,t)\;
{\delta \over \ln (1+\delta)}\;
{ 1\over \sqrt{1+\delta}}
 = T(0,t)\;\left(1 + {\delta^2 \over 24} + {\cal O} \left[\left({m\over 
M}\right)^3\right]\;
\right)\ee
i.e. both definition of temperatures coincide at order $m/M$ on the
surface of the piston.
If we identify $T_B(0,t)$ with the thermodynamic
 temperature then  (\ref{esg}) is the well known expression of the Second Law and 
the  entropy
production $I^{\pm}(t)$ on the right/left of the piston
is strictly zero. 
Thus we must conclude that $i(x,t)=0$ and the increase of the total 
 entropy is entirely due to heat transfer.

\vskip 1mm
On the other hand taking the ``thermal temperature" $T(0,t)$, Eq.~(\ref{59}),  as
the thermodynamic temperature, we 
obtain for the entropy production:
\bea
I^{\pm}&=& {d\over dt} S_B^{\pm}(t)\;
\mp\;
\;{1\over
T^{\pm}(0,t)}\;A\;j_Q^{\pm}(0,t)
\nonumber\\
&&\nonumber\\
&=& \pm\;\left[{\ln(1+\delta)\over
\delta}\;\sqrt{1+\delta}-1\right]\;{A\over
T^{\pm}(0,t)}\;j_Q(0,t)
\nonumber\\
&&\nonumber\\
&=&\mp{[\delta^{\pm}]^2\over 24}\;{A\over T^{\pm}(0,t)}\;j_Q(0,t)
+{\cal O}\left[
\left({m\over M}\right)^4\right]\label{e70}\eea Since $j_Q$ is  order of $m/M$, 
we thus conclude that $i(x,t)$ is zero to second order in $m/M$;
but $j_Q(0,t)$ being negative, the definitions $s_B (x,t)$ and
$T(0,t)$ are not compatible at order 
${\cal O}[(m/M)^3]$ since they lead to an
entropy production which will be negative for some $(x,t)$.

\vskip 3mm
\noindent
\underline{Conclusion}

The distribution functions at the surface of the piston 
$\rho^{\pm}(0,v,t)$, Eqs.~(\ref{eq43}) and (\ref{eq47}),
yield thermodynamic properties which are consistent to first
order in $m/M$ with previous results and with the  laws of
thermodynamics. In particular, the forces on both sides of the
piston are equal, the ``heat flux" $j_Q$ is continuous, the different
definitions of ``temperature" and ``entropy" coincide at this order 
on the surface of the piston. Moreover the viscous stress $\tau + p $ is zero.
If we adopt the Boltzmann's definition
of entropy, then the total entropy
production:
\bea
I(t)&=&{dS_B\over dt}={dS^-_B\over dt}+{dS^+_B\over dt}
\nonumber\\
&&\nonumber\\
&=&{mA\over 2M}\sqrt{8k_B\over m\pi}
(p_0^++p_0^-)\left( \frac{\sqrt{T_0^+}+\sqrt{T_0^-}}{T_0^- T_0^+} 
\right)
\left(\sqrt{T_0^+}-\sqrt{T_0^-}\right)^2
+{\cal O}\left[
\left({m\over M}\right)^2\right]\eea
is  entirely due to heat transfer.
However with this definition of entropy, we are forced to take
$T_B$ as definition of temperature, at least if we want to go
at order $(m/M)^2$.

\section{Propagation through the fluid: similarity flow}

Since the effect of purely elastic collisions between the particles is identical to
the non-interating case,
we consider that the distribution function
$\rho(x,v,t)$ for the gas is the solution of 
the continuity equation (\ref{eq25}), given by:
\be\label{eq60}
\rho(x,v,t)=\rho^-(x,v,t)+\rho^+(x,v,t)\ee
with:
\be
\left\{\ba{l}
\rho^-(x,v,t)=\theta(-x)[
\theta(vt-x)\rho^-(v)+\theta(x-vt)\widetilde{\rho}^-(v)]
\\
\\
\rho^+(x,v,t)=\theta(x)[
\theta(x-vt)\rho^+(v)+\theta(vt-x)\widetilde{\rho}^+(v)]
\ea
\right.\ee
where
\be
\ba{lcl}
\rho(v)=n_0\varphi(v)&{\rm with}&
\displaystyle\varphi(v)=\sqrt{\beta\over \pi}\;e^{-\beta
v^2}\\
&\\
\widetilde{\rho}(v)=n_0\widetilde{\varphi}(v)&{\rm with}&\displaystyle
\widetilde{\varphi}(v)={1\over 1+\delta}\;\sqrt{\beta\over \pi}
\;e^{-\widetilde{\beta}
v^2}\\
&\\
\displaystyle\beta={m\over 2k_BT_0}&{\rm and}  &\widetilde{\beta}=\beta(1+\delta)^{-1}
\ea
\ee
with $\delta$ given by Eq.~(\ref{eq46}) and the subscripts $(+/-)$
corresponding to $(x>0) $ resp. $ (x<0)$ have been omitted.
We thus have a similarity flow, i.e. for all field $f(x,t)$
we have:
\be
f(x,t)=f(\xi)\hskip 10mm {\rm where}\hskip 10mm
\xi={x\over t}\ee
Notice that $\xi$ is negative on the left 
and positive on the right of the piston.

From the expression $\rho(x,v,t)$, Eq.~(\ref{eq60}),
we can compute the thermodynamical quantities defined in Section~3.
We shall detail the computations 
only for the left side of the piston  ($\xi\leq 0  $ and $\delta^- > 0  $)
and shall drop the $(-)$ subscript. The different fields are
represented on Fig.~(1-5) for special values of $n_0^{\pm}$
and  $T_0^{\pm}$. From the definition of density $n(x,t)=n(\xi)$:
\be
n(\xi)=\int_{\xi}^{+\infty}dv \rho(v)+\int^{\xi}_{-\infty}
dv \widetilde{\rho}(v)\ee
follows that:
\be\label{81}
\partial_{\xi}n(\xi)=\widetilde{\rho}(\xi)-\rho(\xi)\ee
Defining $\bar{\xi}$  (in fact $\bar{\xi}^-$ but we have
omitted the subscript $(-)$ in this whole section) by
$\widetilde{\rho}(\bar{\xi})=\rho(\bar{\xi})$, which gives
 (note that $\bar{\xi}<0$):
\be \beta \bar{\xi}^2=(1+\delta)\;{\ln(1+\delta)\over \delta}
\hskip 5mm{\rm hence}
\hskip 5mm\bar{\xi}=-\;\sqrt{2k_BT_0^-\over m}+{\cal O}
\left({m\over M}\right)\ee
we have
\be
\ba{ll}
\widetilde{\rho}(\xi)<\rho(\xi)&{\rm for}\;\;\;\bar{\xi}<\xi\leq 0\\
&\\
\widetilde{\rho}(\xi)>\rho(\xi)&{\rm
for}\;\;\;\xi<\bar{\xi}
\ea
\ee
and $n(\xi)$ reaches a maximum at
$\xi=\bar{\xi}$  (see Fig.~1).
For the velocity field we have (Fig.~2):
\be\label{84} 
n(\xi)w(\xi)=
n_0\sqrt{\beta\over\pi}
\;{1\over 2\beta}\;
\left(e^{-\beta\xi^2}-e^{-\widetilde{\beta}\xi^2}
\right)
\ee
Therefore, $w(\xi)$ is negative for $\xi<0$, zero for $\xi=0$
and is minimum for $\xi=\bar{\xi}$.
For the stress tensor  $\tau(x,t)=-2u(x,t)$ we have  (Fig.~3) :
\be
-\,\partial_{\xi}\tau(\xi)=m[\widetilde{\rho}(\xi)-
\rho(\xi)]\;[\xi-w(\xi)]^2\ee
i.e. the only points where $\partial_{\xi}\tau=0$ are
$\xi=0$ and $\xi=\bar{\xi}$. For the heat flux we have  (Fig.~4) :
\be\label{jQ}
j_Q(\xi)={mn_0\over 2}\;\sqrt{\beta\over \pi}\;
{1\over 2\beta^2}\;\left\{
\left[e^{-\beta\xi^2}-e^{-\widetilde{\beta}\xi^2}
\right]\;\left[1+\beta\xi^2-{3\over 2}{T(\xi)\over T_0}
\right]
-\delta e^{-\widetilde{\beta}\xi^2}
\right\} -{m\over 2}n(\xi)w(\xi)^3\ee
\be
\partial_{\xi}j_Q(\xi)=
{1\over 2}[\widetilde{\rho}(\xi)-
\rho(\xi)]\;[\xi-w(\xi)]
\;\left[m[\xi-w(\xi)]^2-6{u(\xi)\over n(\xi)}\right]
\ee
and thus $\partial_{\xi}j_Q=0$  for $\xi=0$,
$\xi=\bar{\xi}$ and $\xi=\widehat{\xi}^-$
where:
\be\label{85}
\widehat{\xi}^-=-\;\sqrt{3k_BT_0^-\over m}+{\cal O}
\left({m\over M}\right)\ee
If we adopt the thermal definition of temperature, i.e.
$k_BT=2u/n$,  we find that  (Fig.~5 a) :
\be\label{88}
\partial_{\xi}[k_BT(\xi)]={1\over n(\xi)}\;
[\widetilde{\rho}(\xi)-
\rho(\xi)]\;\Big{\{}m[\xi-w(\xi)]^2-k_BT(\xi)\Big{\}}
\ee
and thus $\partial_{\xi}T=0$ if $\xi=\bar{\xi}$ and
$\xi=\widetilde{\xi}$ where:
\be
\widetilde{\xi}=-\;\sqrt{k_BT_0^-\over m}+{\cal O}
\left({m\over M}\right)\ee
Therefore $\partial_{\xi}T$
is negative  for $\xi=x/t\in\;]\bar{\xi},\widetilde{\xi}[$
and positive if $\xi\not\in\;]\bar{\xi},\widetilde{\xi}[$(Fig.~5 b).

In conclusion with the thermal definition of temperature, the contribution
$j_Q\partial_{\xi}(1/T)$ to the irreversibility is non-negative
if $\xi\not\in\;]\bar{\xi},\widetilde{\xi}[$ but negative 
  for $\xi=x/t\in\;]\bar{\xi},\widetilde{\xi}[$ .
Therefore, whatever is the definition of entropy and of $j_S$, it is impossible to obtain
an equation of the form
\be
\partial_ts+\partial_x(sw+j_S)=i\ee
with  $i=j_Q\partial_{\xi}(1/T)$
without violating the Second Law (Fig.~6 a).
Of course if we take Boltzmann's definition, Eq.~(\ref{29}), 
and $j_S=j_B$, then  $i(x,t)=0$.


\section{Entropy and entropy production }

Whatever is the definition of entropy, from the Second Law
\be 
\partial_t s+ \partial_x (sw+j_S) = i, \qquad i(x,t) \geq 0\nonumber
\ee
and the similarity of the flow, i.e.\\
 \be s(x,t)= s(\xi), \quad t .i(x,t)= t. i(\xi), \quad \xi=\frac{x}{t}\ee
 it follows
that 
\be\label{e1}
t.i(x,t)=t.i(\xi)= (w-\xi) \partial_{\xi}s +s\partial_{\xi}w +\partial_{\xi}j_S
\ee
and
\bea
\frac{d}{dt}S^-(t)&=&A\,\,\left\{  \int_{-\infty}^0\,\,\,d\xi \,\,t .i(\xi) - 
   j_S(\xi=0)         \right\}\nonumber\\
&=&I^-(t) -A\,\,\ j_S(\xi=0)    
\eea
If we take Boltzmann's definition of entropy,  
\bea
s_B^-(x,t)=s_B^-(\xi)&=&
\theta[-\xi]\,\bigg{\{  }  s_B^-(-\infty) + k_B   \int_{-\infty}^{\xi}\,\,dv \,\,[\rho^-\ln{\rho^-} -
\widetilde{\rho}^-\ln{\widetilde{\rho}^-}]  \bigg{ \}} \nonumber\\
s_B^-(-\infty)&=&n_0^-\,k_B\,\frac12\,\left[1+ \ln{\frac{2\pi k_B T_0^-}{m(n_0^-)^2}}\right]
\eea
then, as mentionned in section~3, if we define the entropy current by Eq. 31, i.e. $j_S=j_B$, we will have
$i(x,t)=0$ and to obtain the Second Law in the usual form (with $I^-(t)=0$), i.e.,
\bea
\frac{d}{dt}S^-_B&=&\frac{1}{T(0,t)}\,\,P_Q^{(+)\to(-)}(t)
\eea
we have to introduce a ``Boltzmann temperature" defined by 
\bea
T_B(x,t)&=&\frac{j_Q(x,t)}{j_B(x,t)}
\eea
However  with such a definition we do not recover the usual phenomenological equations of
thermodynamics.\ \\
On the other hand if we take Boltzmann's  definition of entropy and define the entropy current by
the usual phenomenological equation $j_Q=T j_S$, where $T= 2u/nk_B$ is the thermal temperature, then
$i(x,t) \neq 0$ given by Eq.~(\ref{e1}).
 However from Eqs.~(\ref{e68})  and (\ref{e70})
\bea
I^-(t) = A \frac{j_Q(0,t)}{T(0,t)}\,\,\left(
 1- \sqrt{1+\delta^-} \,\frac{\ln(1+\delta^-)}{\delta^-}\right)
\eea
is negative. In other words, as already noticed in Section~6,  $i(x,t)$ can not be non-negative (see Fig.~6 a) which is not compatible with
the Second Law. 

If we adopt the definition of entropy given by the local equilibrium condition of a
simple one-component fluid, i.e.
\bea
s(\xi) &=& n_0(\xi)\,\,k_B\frac12 \left\{
 1+ \ln{\frac{4\pi\,u(\xi)}{m n^3(\xi)}} \right\}\eea
then \bea
T(\xi) = \frac{2 u(\xi)}{k_B n(\xi)} \quad {\rm and}
 \quad i(\xi) =  j_Q \partial_{\xi} \left(\frac{1}{T}\right)\eea
As discussed in Section~6 this leads again 
to the conclusion that the Second Law is violated.

\section{Two-component fluid description}

To show that the time evolution of the piston problem satisfies the laws 
of thermodynamics we first remark that the fluid on the left side is
 characterized by 3 parameters $(n_0^-, T_0^-, T_P$). This 
leads us to look at our problem as a two-component fluid. \vspace{0.2cm}
\\
 At the surface $x = 0$, introducing 
\be
n_1(0,t)={n_0\over 2}, \hskip 10mm
n_2(0,t)={n_0\over 2\sqrt{1+\delta}}, \hskip 10mm
n=n_1+n_2\ee the Boltzmann entropy Eq.~(\ref{eq53}) can be express as
\bea
s_B(0,t)&= &s(u(0,t),n_1(0,t),n_2(0,t))\\
s(u,n_1,n_2)&=& n\,\, k_B\left\{    \frac12 (1+\ln(4\pi) )+\frac12 \ln{\frac{u}{(n_1+n_2)^3} }+g(n_1,n_2) 
 \right\}\label{su2}
\eea \vspace{ -0.25cm}
and\vspace{ -0.3cm}
\be\label{eq54-2}
g(n_1,n_2)\equiv C(\delta)=2\,\left(\frac{n_2}{n_1+n_2}-{3\over
4}\right)\,\ln{n_1\over n_2}
+\ln\left(1+{n_1\over n_2}\right)-\ln 2\label{gg}\ee
We note that $g(n_1,n_2)=g(n_1/n_2)$ implies 
\be
n_1{\partial g\over \partial n_1}
+n_2{\partial g\over \partial n_2}=0\ee
From (\ref{su2}) we obtain for the temperature
$T_s=\partial u/\partial s$ and  the pressure $p$,  Eq.~(\ref{eq8}),  at the surface
of the piston:
\be\label{eq55-2}
\left\{
\ba{l}\displaystyle
T_s(0,t)=2\;{u(0,t)\over k_Bn(0,t)}=T(0,t)=T_0\sqrt{1+\delta}\\
\\\displaystyle
p(0,t)=n(0,t)\,k_BT(0,t), \;\;\;{\rm i.e.}\;\;
\tau_{fr}=\tau+p=0
\ea
\right.\ee
where the subscript $(+/-)$ for right/left have been
here omitted.

\vskip 3mm
\noindent
In the bulk ($x \neq 0$) we shall assume  
that for all $\xi$ one can express
the variables $\{n_0, T_0, T_P\}$ in function of $\{u,n_1,n_2\}$,
$n=n_1+n_2$ as was the case for $x = 0$. The thermal temperature will then be of the form
\be
T(\xi)\equiv {2u\over nk_B}=T(\xi,u,n_1,n_2) \label{termT}\ee
and thus if we want to identify the temperature Eq.~(\ref{termT}) with the thermodynamic temperature
$T_s={\partial u}/{\partial s}$ the thermodynamic entropy must be given by
\bea \label{ss}
s(\xi)&=& s(\xi,u,n_1,n_2)\nonumber\\
&&\nonumber\\
&=&
n(\xi)k_B\left\{
{1\over 2}\left(1+\ln {4\pi\over m}\right)+{1\over 2}
\ln {u\over n^3}+g(\xi,n_1,n_2)
\right\}
\eea
where at this point $g(\xi,n_1,n_2)$ is undetermined.
For simplicity we choose $g$ independent of $\xi$, thus given by 
Eq.~(\ref{gg}), i.e.
\be
 g(n_1,n_2)=2\,\left({n_2\over {n_1+n_2}}-{3\over
4}\right)\,\ln{n_1\over n_2}
+\ln\left(1+{n_1\over n_2}\right)-\ln 2\ee
Note that if $n_1(\xi)$ is chosen such that  $n_1(\xi=\pm \infty) =\frac12 n_0^{\pm}$ then we will
have $s=s_B$ for $\xi=\pm \infty$ and for $\xi=\pm\,\, 0$, i.e. at those points where the fluids are in a stationary state.
Moreover since  $g( n_1, n_2)=g( n_1/ n_2)$ we have 
$n_1\partial g/\partial n_1+ n_2\partial g/\partial n_2=0$
which
implies 
\be
p=nk_BT,\quad {\rm i.e.} \quad \tau_{fr}=\tau+p=0.\nonumber\ee
In that case it follows that
\be
i=j_Q\partial_x\left({1\over T}\right)
+j_1\partial_x\left({\mu_2-\mu_1\over T}\right)\ee
\bea
\mu_2-\mu_1&=& k_BT \left( 2\ln{\frac{n_2}{n_1}}+ \frac12 \left(\frac{n_2}{n_1}-\frac{n_1}{n_2}  \right)\right)
={\cal O}\left({m\over M}\right)
\eea
which yields
\be\label{eqti}
t.i(\xi)=-{1\over T_0^2}j_Q\partial_{\xi}T+
6k_B\;{j_1\over n_0}\;(\partial_{\xi}n-2\partial_{\xi}n_1)
+{\cal O}\left[\left({m\over M}\right)^3\right]
\ee
Moreover, the continuity equation for $n_1$
\be\label{eq103}
-\xi\partial_{\xi}n_1+\partial_{\xi}(n_1w+j_1)=0\ee
together with the boundary conditions
\be
\lim_{\xi\to -\infty}j_1(\xi)=0, \hskip 10mm
\lim_{\xi\to -\infty}n_1(\xi)={n_0\over 2}\ee
implies (for $\xi < 0$)
\be
j_1(\xi)=\xi(n_1-{n_0\over 2})-n_1w-\int_{-\infty}^{\xi}
d\xi\left(n_1-{n_0\over 2}\right)\ee
Let us then look for a solution of the form
\be\label{101}
n_1(\xi)={n_0\over 2}\;\left(1+{\lambda_1\over \bar{\xi}}\,w + \lambda_2\,\, \partial_{\xi} w\right)\ee
where  at this  stage $\lambda_1$ and $\lambda_2$ are unspecified constants.
With this ansatz, and introducing\\  $x = -\sqrt{\beta}\xi > 0$, we have for the left compartment
\be\label{112}
j_1(\xi)=\delta\,n_0 \sqrt{\frac{\beta}{\pi}}  \frac{e^{-x^2}}{4\beta}
\left(x^2+\lambda_2 x^2 (2x^2-1) -\lambda_1\left(x^3+\frac12 x+\frac12 e^{x^2}  \int_x^{\infty} du\,\,
e^{-u^2}  \right)      \right)+
{\cal O}\left[\left({m\over M}\right)^2\right]
\ee
and 
\be\label{113}
\partial_{\xi} (n-2n_1)= \delta\, n_0 \sqrt{\frac{\beta}{\pi}}\, e^{-x^2} 
\big\{ (x^2-1)\,[     1 -    \lambda_1 x   +   \lambda_2  (2x^2-1) ]-2\lambda_2 x^2  
       \big\}+
{\cal O}\left[\left({m\over M}\right)^2\right]
\ee
We thus have to find $\lambda_1$  and $\lambda_2$ such that $ t\,i(\xi)$, 
Eq.~(\ref{eqti}), is non negative,
where
\be\label{1112}
j_Q(\xi)\partial_{\xi}\frac1T= \delta^2\, n_0 k_B\, {\frac{ e^{-2 x^2} }{4\pi}}\,  (x^2-1)(2x^2-1)[2+x^2(2x^2-1)]+
{\cal O}\left[\left({m\over M}\right)^3\right]
\ee

Moreover, we will have 
\bea
\frac{d}{dt}S^-(t)&=&A\,\,  \int_{-\infty}^0\,\,\,d\xi \,\,t i(\xi) -   \frac{A}{T(-0,t)}\,[j_Q(0,t)
+j_1(\mu_2-\mu_1)(-0,t)]        
\eea
with
\bea\label{122}
j_Q(0,t)+ j_1(\mu_2-\mu_1)(-0,t)&=& j_H(-0,t)\nonumber\\
&=&j_Q(0,t)\left\{1- \frac18\lambda_1 \sqrt{\pi} \left[\ln{(1+\delta^-)}
+\frac12 \frac{\delta^-}{\sqrt{1+\delta^-}}\right]\right\}\nonumber\\
&=&j_Q(0,t)\left[1- \delta^-\frac3{16}\lambda_1
 \sqrt{\pi} \right] +{\cal O}\left[\left({m\over M}\right)^3\right]
\eea and $j_Q(0)$ is given by Eq.~(\ref{eq52}).
Using numerical computation, we see that choosing 
$
\lambda_1 =1 $ and $ \lambda_2=0.1 
$
the entropy production Eq.~(\ref{eqti}) will be strictly positive (Fig.~6 b). For the right compartment $(\xi > 0)$ we have
\bea \label{etoile}
j_1(\xi)&=&\xi\left(n_1(\xi)-{n_0\over 2}\right)-n_1(\xi)w(\xi)+\int_{\xi}^{\infty}
d\xi\left(n_1(\xi)-{n_0\over 2}\right) \qquad\quad(\xi > 0)\\
\frac{d}{dt}S^+(t)&=&A\,\,  \int_0^{\infty}\,\,\,d\xi \,\,t i(\xi) +  \frac{A}{T(+0,t)}\,[j_Q(0,t)
+j_1(\mu_2-\mu_1)(+0,t)]    \eea 
\bea
j_Q(0,t)+ j_1(\mu_2-\mu_1)(+0,t)= 
j_Q(0,t)\left[1- \delta^+\frac3{16}\lambda_1
 \sqrt{\pi} \right] +{\cal O}\left[\left({m\over M}\right)^3\right]
\eea and
 Eq.~(\ref{112}-\ref{1112}) will remain valid with $x =  \sqrt{\beta^+} \xi $. Therefore with  $
\lambda_1 =1 $ and $ \lambda_2 = 0.1 
$  the entropy production $ i(x,t)$ is strictly positive for all $(x,t)$, and all phenomenological
equations  of thermodynamics will hold (Fig.~6 a).
\  \\

\noindent\underline{Remarks}
\begin{enumerate}

\item With the choice $\lambda_1 =1 $ and $ \lambda_2=0.1
$ the density $n_1(x,t)$ corresponds approximately  to the density $n_w(x,t)$ of particles  with
velocity larger than the average velocity of the fluid $w(x,t)$ (see Fig.~7).

\item  We note that both definitions of heat current, (see Section~2 ) $j_Q$ or $j_H = j_Q + j_1
(\mu_2-\mu_1)$ are quantities of order $ (m/M$), which
differ by a factor of order $(m/M)^2$ (see Eqs.~(\ref{112}-\ref{113}), or
 (\ref{122})). 
However $j_Q$ is continuous at the surface of the piston (as it should since it is the energy current), 
while $j_H$ is not continous (see Fig.~9).

\item From Eqs.~(\ref{112}-\ref{113})
 follows that the possible choice of $\lambda_1$, $\lambda_2$, leading to a positive entropy production, 
will be independant  of the choice of the initial conditions ($T_0^{\pm}, n_0^{\pm}$). In all the numerical 
computations (Figs.~1-9) the following parameters have been chosen
\bea
m = 1 \qquad\quad \qquad M = 10.000\qquad \qquad\quad k_B = 1 \nonumber \\
T_0^- = 1\qquad   T_0^+ = 10\qquad n_0^- = 3000\qquad   n_0^+ = 300.085\nonumber \\
{\rm i.e.}\qquad  p_0^- = 3000\qquad\qquad   p_0^+ = 3000.85
\eea
\item Let us note that the difference between the entropy densities $s(x,t)$ and $s_B(x,t)$ is very 
small  of order $(m/M)^2 $ (fig. 8 b).
\item We recall that for one-dimensional ideal fluid the velocity of sound is $\sqrt{\frac{3k_B T}{m}}$
which coincide with $\widehat{\xi}^-$, Eq.~(ref{85}).
\item If we consider the stationary states of our model, then the left compartment is 
an homogeneous fluid in a stationary state caracterised macroscopically by its 
temperature $T$ (i.e.  $T^-$) its density $n$
 (i.e. $ n^-$), its velocity $w(x,t) =0$ and the heat flux $j_Q$. Microscopically this fluid 
is described by the density function
\bea
\rho(x,v;t)& = &\theta[-x]\,\,\bigg[\, \theta[v]\,\, 2 n_1\,\, \sqrt{\frac{\beta}{\pi}} \,\,e^{-\beta v^2} + \,\,\theta[-v]\,\, 2 n_2\,\, \sqrt{\frac{\widetilde{\beta}}{\pi}} \,\,e^{-\widetilde{\beta} v^2}\bigg]
\eea where, using the previous results, \be\label{bet}
\left\{\ba{l}\displaystyle{\beta} =
\displaystyle{\frac{m}{2k_BT}\,\,\frac{n_1}{ n_2 }} \qquad\qquad \qquad\qquad\qquad\quad \,\,\,\,\widetilde{\beta} = \displaystyle{\frac{m}{2k_BT}\,\,\frac{n_2}{ n_1 }}\\
\displaystyle{j_Q = - k_BT \left(\frac{n_1}{ n_2 }-\frac{n_2}{ n_1 }\right) \,\,\sqrt{\frac{n_2}{ n_1 }}\,\sqrt{\frac{2k_B}{m \pi}};\qquad n = n_1 + n_2}
\\
\ea\right.\ee
and the temperature of the piston is
\bea
T_P & =  & T\,\,\left\{ \frac{n_2}{ n_1 }  + \frac{(M + m)^2}{4 m M}\,\,\left(\frac{n_1}{ n_2 }-\frac{n_2}{ n_1
}\right)
\right\}
 \eea
The system is furthermore caracterised by its entopy function $s = s(u, n_1, n_2 )$, given
 by Eq.~(\ref{ss}) and we have: \be\label{jjjk}
\left\{\ba{l} \displaystyle {u =\frac12\,\,n \,\,k_BT            }\qquad \displaystyle{p =\,\,n \,\,k_BT  
 = - \tau        }\\
\\
{\displaystyle s[u(x,t), n_1(x,t), n_2(x,t) ] = s_B(x,t)}
\\
\ea\right.\ee
Moreover in this stationary state $j_1 = 0 $ implies 

\bea
\frac{d}{dt}S^-(t)&=& -A\,\, \frac{j_Q}{T} \qquad\qquad I = 0\eea
We obtain similar expressions for the right compartment and the stationary condition of the state is expressed by the relations 
\bea
p^- = p^+ \qquad\quad T_P^- = T_P^+\qquad \qquad j_Q ^- = j_Q^+
\eea

Finally the total entropy production in the stationary state of this isolated system
\bea
\frac{d}{dt}S= -A\,j_Q\,\left( \frac{1}{T^-} - \frac{1}{T^+}\right)
\eea is entirely due to the heat transfer from the subsystem at temperature  
$T^+$ to the subsystem at temperature $T^-$which reflects the fact that the fluids
 are ideal. (Let us note that   $S= \infty$ and thus only the entropy production 
is relevant in such non-equilibrium stationary states.)

We should stress that the fluid in the stationary state  is not locally in equilibrium 
(see Eq.~(\ref{etoile})), which is a general property of far-from-equilibrium stationary states, 
but is a two-component simple fluid.

\end{enumerate}

\section{Conclusions and open problems}

Taking initial conditions such that on the average the piston remains at the same point so that there is only heat transfer from the right to the left compartment ($ T_0^+> T_0^-$), but  no work done, we have first derived the time evolution of the fluids on both sides of the piston.


We have then shown that this fluid 
  out of equilibrium can be considered 
 as a 2-component  simple fluid 
 described by an entropy function $s = s(u,n_1,n_2)$, and all the properties of the fluid 
will obey the laws of thermodynamics. In particular the two laws of thermodynamics are satisfied with
\bea
e&=&{1\over 2}\;mnw^2+u(s,n_1,n_2) \nonumber \\
T &=& \frac{\partial u}{\partial s} = \frac{2u}{n k_B}\nonumber \\
p &=& s T + n_1\mu_1+ n_2\mu_2 -u = nk_BT = -\tau \nonumber \\
j_e &=& -\tau w +j_Q
\eea
where $n, w, u, \tau, j_Q$  are defined by the usual relations of kinetic theory, $ \mu_i =  \partial u/\partial n_i$ 
\bea
T j_S = j_Q + j_1 (\mu_2-\mu_1)
\eea
 and
 \bea
i =  j_Q \partial_x\left(\frac1T\right) + j_1  \partial_x\left(\frac{\mu_2-\mu_1}{T}\right)
\eea
 is strictly positive.
 
 Moreover the entropy function which we have introduced coincides
 with the Boltzmann entropy
 in those domains where the fluids are in a stationary state. It was also shown that other natural 
definitions of entropy, temperature, irreversibility, which one could consider led to results violating 
the Second Law. Since the evolution is a similarity flow, one can not expect to obtain phenomenological 
relations of the form
 \bea
 j_Q = C_{11}\,\,\partial_x\left(\frac1T\right) +  C_{12} \,\,\partial_x\left(\frac{\mu_2-\mu_1}{T}\right)\nonumber \\
 j_1 = C_{21}\,\,\partial_x\left(\frac1T\right) +  C_{22}\,\, \partial_x\left(\frac{\mu_2-\mu_1}{T}\right)
\eea
 with $C_{ij}$ a state function (indeed ${\displaystyle\partial_x = \frac1t \partial_{\xi}}$).
 
 However usual relations such as 
\bea
\frac{d}{dt}S^- (t) &=&I^-(t)  + \,\,\frac{P_Q^{(+)\to(-)}}{T^-(0,t)}\nonumber \\
\frac{d}{dt}S^+ (t) &=&I^+(t)  - \,\,\frac{P_Q^{(+)\to(-)}}{T^+(0,t)}
\eea
 with  
\be I^{\pm}(t)\,\,\, > \,\,\,0\ee
and 
\be
P_Q^{(+)\to(-)}\,\, =\, \,\kappa (T^+_0 - T^-_0)
\ee
are satisfied with 
 \bea
\kappa = A\,\,\frac{m}M \sqrt{\frac{8 k_B}{m\pi}}\,\,\frac12 \,\,\left(\frac{p^-_0 + p^+_0}{T^-_0 + T^+_0}\right)
+{\cal O}\left[\left({m\over M}\right)^2\right]\quad > 0
\eea
It is however an open problem to understand what is 
the physical meaning of the decomposition of the density
 as a sum of two densities. It is also 
to be understood whether there is a  physical argument which would lead to a uniquely
 defined  decomposition, and whether one could find a microscopical definition of our 
entropy function, i.e. whether it is possible to find a state function $s(x,v;t)=s\big(\rho(x,v;t)\big)$ such that 
\bea
s(x,t) = s\left[u(x,t), n_1(x,t), n_2(x,t) \right]=\int\,\,dv \,\,s(x,v;t)
\eea

\ \\ \\
{\bf Acknowledgements:}

A. Lesne greatly acknowledges the hospitality at the
Institute of Theoretical Physics of the \'Ecole Polytechnique F\'ed\'erale
de Lausanne where part of this research has been performed.
 We also thank the ``Fonds National Suisse de la Recherche Scientifique''
for his financial support of this project.

\newpage
\noindent
{\Large\bf Figure captions}

\vskip 15mm
\noindent
{\bf Figure 1:}  Density of particles $n(x,t) = n(\xi)$
\vskip 5mm
\noindent
{\bf Figure 2:}  Fluid velocity  $w(x,t) = w(\xi)$
\vskip 5mm
\noindent
{\bf Figure 3:} Pressure $p(x,t) = p(\xi) = -\tau(\xi)$
\vskip 5mm
\noindent
{\bf Figure 4:} Heat current $ j_Q(x,t) =  j_Q(\xi) $

\vskip 5mm
\noindent
{\bf Figure 5:}

a) Temperature $T(x,t) = T(\xi)$

b) Gradient $\partial_{\xi} T(\xi)$

\vskip 5mm
\noindent
{\bf Figure 6:} Entropy production 

a)  Contribution $ j_Q  \,\, \partial_{\xi}(1/ T) $

b) $i(x,t) =t\, i(\xi)$

\vskip 5mm
\noindent
{\bf Figure 7:}   Density  $n_1(x,t)/n_0 = n_1(\xi)/n_0$ and \\ density of particles  $n_w(x,t)/n_0 = n_w(\xi)/n_0$  with velocity larger than $  w(x,t)$ 

\vskip 5mm
\noindent
{\bf Figure 8:} 

a) Entropy density $s^{\pm}(x,t)/s^{\pm}(\pm \infty)  $

b) $[ s(x,t) - s_B (x,t)]/s^{\pm}(\pm \infty)$

\vskip 5mm
\noindent
{\bf Figure 9:}
Contribution $ [j_1  (\mu_2 -\mu_1) ](x,t) = [j_1  (\mu_2 -\mu_1) ](\xi)$ to the entropy current. \\Note that it is of the order  $ 10^{-4} \cdot j_Q(x,t) $

\end{document}